\documentclass[preprint,12pt]{elsarticle}

\usepackage{amsmath,epsfig}
\usepackage{amssymb,amsfonts,amsthm,bm}
\usepackage{pstricks}

\newrgbcolor{green}{0 1 0}
\newrgbcolor{darkgreen}{0 .65 0}

\newcommand{\GARCH}{{\small GARCH}}

\newcommand{\vol}{{\sc vol.}}
\newcommand{\no}{{\sc no.}}

\journal{Digital Investigation}

\begin{document}

\begin{frontmatter}



\title {A Novel Contourlet Domain Watermark Detector for Copyright Protection}

\author{Maryam Amirmazlaghani\\
}

\address{ {\small Amirkabir University of Technology, Department of Computer Engineering and Information Technology}\\
{\small P.O. Box 15914, Tehran, Iran}\\
{\tt mazlaghani@aut.ac.ir}\\
{\small Tel: 98 21 64542704, Fax: 98 21 66406469}}

\def\x{{\mathbf x}}
\def\L{{\cal L}}


\begin{abstract}
\bf Digital media can be distributed via Internet easily, so, media owners are eagerly seeking methods to protect their rights. A typical solution is digital watermarking for copyright protection. In this paper, we propose a novel contourlet domain image watermarking scheme for copyright protection. In the embedding phase, we insert the watermark into the image using an additive contourlet domain spread spectrum approach. In the detection phase, we design a detector using likelihood ratio test (LRT). Since the performance of
the LRT detector is completely dependent on the accuracy of the
employed statistical model, we first study the statistical
properties of the contourlet coefficients. This study demonstrates
the heteroscedasticity and  heavy-tailed marginal distribution of
these coefficients. Therefore, we propose using two dimensional
generalized autoregressive conditional heteroscedasticity
(2D-GARCH) model that is compatible with the contourlet
coefficients. Motivated by the
modeling results, we design a new watermark detector based on
2D-GARCH model. Also, we analyze its performance by computing the
receiver operating characteristics. Experimental results confirm
the high efficiency of the proposed detector. Since a watermark detector for copyright protection should be robust against attacks, we examine the robustness of the proposed detector under different kinds of attacks.\\
\end{abstract}
\begin{keyword}
Image Watermarking, Copyright Protection, Watermark Detector,
, Contourlet Transform, 2D-GARCH model.
\end{keyword}
\end{frontmatter}
\section{INTRODUCTION}
The Internet is an efficient distribution system for digital
media. Data distribution on the internet increases the importance
of the data security and copyright protection issues. A typical
solution is digital watermarking. Digital watermarking can be
defined as the practice of imperceptibly altering a media content
to embed a message about that media.   Digital watermarks can be
applied to different media contents such as image, video, and
audio. Also, watermarking can be used in a wide variety of
applications such as broadcast monitoring, data authentication,
and copyright protection \cite{item18}. In copyright protection,
the main goal is watermark detection, i.e., it is enough to decide
whether a received media contains a watermark
generated with a certain key \cite{item1,item2}.
In other applications, the watermark decoding may be required, i.e., the watermark serves as a secret message that should be decoded correctly \cite{item33,item3}. In this paper, we focus on image watermarking for copyright protection. \\
In the literature, different watermarking methods for copyright
protection have been proposed. They can be classified based on the
domain in which the watermark is embedded for example pixel or
transform domain. Due to the watermark embedding method, the
watermarking schemes can be classified into two main groups:
quantization based \cite{item6,item7} and spread spectrum based
approaches \cite{item66,item4,item5,item10}. The spread
spectrum watermarking  is so popular because it provides a very
high level of security and robustness. In this scheme, a
pseudorandom signal is added into the original media. Spread
spectrum approaches use a transform domain for watermark
embedding. Different types of transforms such as the discrete
Fourier transform (DFT) \cite{item8}, the discrete cosine
transform (DCT) \cite{item1}, the discrete wavelet transform (DWT)
\cite{item5,item10,item99,item999} and the contourlet transform
\cite{item9999,item888,item144} have been used.
 Contourlet transform is an efficient extension of the wavelet
transform using multiscale and multidirectional filterbanks. This
transform provides nearly critical sampling while permits
different number of directions in each scale \cite{item1c}. From
the viewpoint of watermarking, spreading property of the
contourlet transform is important since embedding the watermark
signal into a specific subband results in spreading out the
watermark signal in all subbands during the reconstruction of the
watermarked image \cite{item888}. Recently, due to the good
properties of the contourlet transform, a number of watermarking
methods have been
proposed in this domain \cite{item9999,item888,item144,item8888,item81,item82,item833}. \\
In frequency domain watermarking, the correlation detector has
been used most commonly  \cite{item833,item83,item9,item11}. This
detector is optimal only when the distribution of data samples is
Gaussian. To achieve an optimal detector for the non-Gaussian
data, Bayesian log-likelihood ratio test (LRT) can be employed.
The choice of the statistical model used in the LRT is of great
importance. Several different priors have been considered for
modeling the frequency domain coefficients in watermark detection
such as Laplacian \cite{item13}, generalized Gaussian
\cite{item14}, Bessel K form \cite{item144} and alpha-stable \cite{item888}.\\ In this paper,
we use contourlet domain for watermarking and it is known that the
contourlet coefficients are non-Gaussian \cite{item888,item84}.
Due to the distribution used for the contourlet coefficients,
different types of LRT detectors can be achieved. Previously
proposed models for the contourlet coefficients assume that these
coefficients are identically distributed \cite{item888,item84}. In
the current work, we demonstrate that this assumption is not
compatible with the contourlet coefficients and these coefficients
are
 heteroscedastic, i.e., their conditional variance is not constant.  So, to overcome
the limitations of the previously proposed watermark detectors, we
suggest employing generalized autoregressive conditional
heteroscedasticity (GARCH) model for the contourlet coefficients.
This model proposed by Bollerslev in \cite{item16} for the
financial time series. 2D-GARCH which is the extension of GARCH
model into two dimensions, has been discussed in
\cite{item12,item122,item12m}. This model allows the conditional variance
to change over two dimensions \cite {item12,item12n}. We show that 2D-GARCH model can
capture the important characteristics of contourlet coefficients
such as heteroscedasticity and heavy-tailed marginal distribution.
This model provides an efficient structure for the intrascale
dependencies of contourlet coefficients. Due to the modeling
results, we design a novel watermark detector in the contourlet
domain based on 2D-GARCH model. Our approach is based on solid
statistical theory and we derive the ROC of the 2D-GARCH based
detector analytically. Experimental results confirm the high
efficiency of
the proposed watermarking method.\\
It should be mentioned that this paper is the first work that considers and captures the heteroscedasticity  of the contourlet coefficients. We think that taking into account the heteroscedasticity of the contourlet coefficients can lead to great results in other fields such as contourlet domain image denoising.\\
This paper is organized as follows. In section 2, we review the
contourlet transform. Section 3 discusses the statistical modeling
of the contourlet coefficients. In this section, we introduce
2D-GARCH model and  study its compatibility with the contourlet
coefficients. Watermark embedding process is explained in section
4.  Section 5 describes the 2D-GARCH based watermark detector and
analyzes its performance. Section 6  Sreports the simulation
results. In this section, the experimental performance of the
proposed detector is evaluated  and compared with two other
related watermark detectors. Finally, section 7 concludes the
paper.
\section{CONTOURLET TRANSFORM}
\begin{figure}
\begin{center}
\includegraphics[scale=.38]{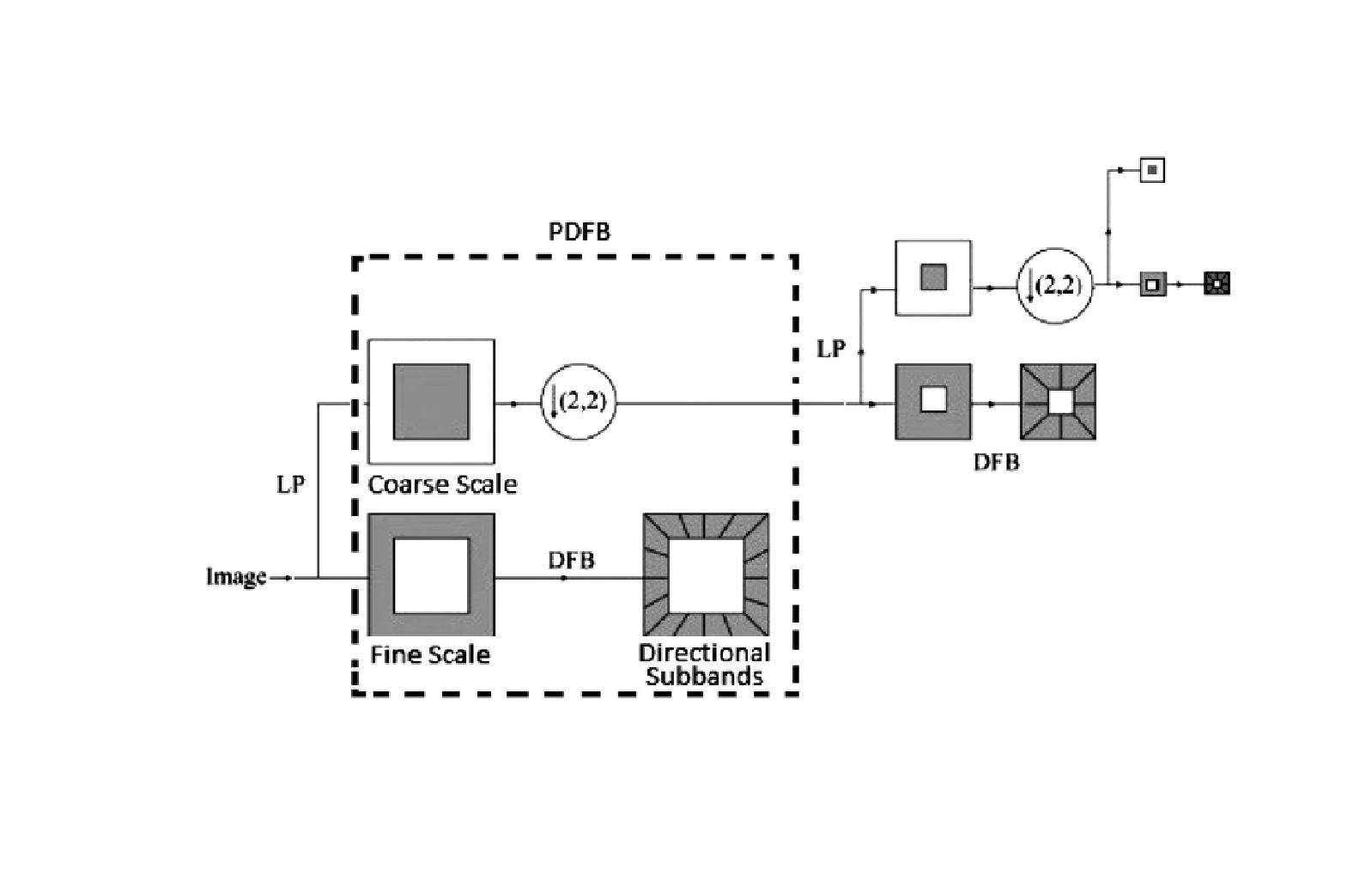}
\end{center}
\caption{\label{fig1} The contourlet transform filter bank}
\end{figure}

\begin{figure}
\begin{center}
\includegraphics[scale=.38]{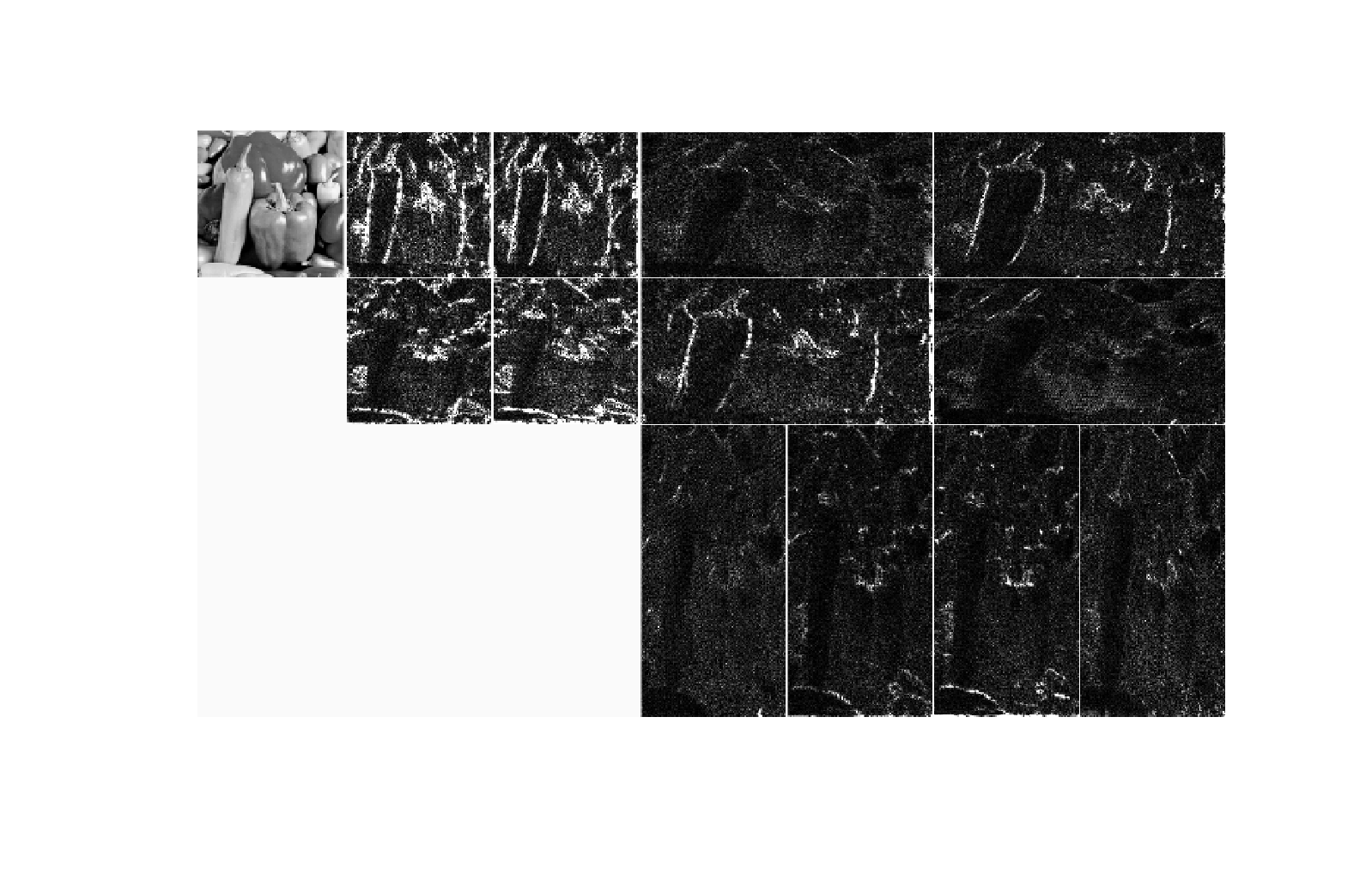}
\end{center}
\caption{\label{fig2} Contourlet transform on the {\it Peppers} image: The contourlet transform with two pyramidal levels is used which followed by four directions in the first level and eight directions in the second level. Large coefficients are
shown in white and small coefficients are shown in black \cite{item1c}} \vspace*{4pt}
\end{figure}

The wavelet transform is an efficient tool for one dimensional
piecewise smooth signals; but, in dealing with two dimensional
signals, it can not  efficiently represent the singularities. So,
to capture the intrinsic geometrical structures in natural images,
many directional image representations have been developed
recently such as dual-tree complex wavelet \cite{itemc2},
ridgelets \cite{itemc3}, curvelets \cite{itemc4}, and contourlets
\cite{item1c}. Contourlet transform provides a sparse expansion
for typical images having smooth contours. This transform consists
of two major stages: the subband decomposition using Laplacian
Pyramid (LP) \cite{itemc5} and  the directional transform using
Directional Filter Banks (DFB) \cite{item1c}. Fig.~\ref{fig1}
represents the relation between LP and DFB decomposition.
Combining  LP and  DFB makes Pyramidal Directional Filter
Bank(PDFB).
Contourlet transform includes one or many PDFB stages. In the
contourlet transform, multiscale and directional decomposition
stages  are independent of each other and different scales can be
decomposed into different numbers of directions. The contourlet
transform provides a high level of flexibility in decomposition
while being close to critically sampled \cite{item1c}. Other
directional representations are significantly overcomplete or
provide a fixed number of directions. The subbands of contourlet
transform for the {\it Peppers} image have been shown in
Fig.~\ref{fig2}. It is clear from this figure that only
contourlets that fit with both location and direction of the image
edges produce significant coefficients.
\section{STATISTICAL MODELING}
It is known that the contourlet coefficients are highly
non-Gaussian \cite{item888,item84}. The histograms of these
coefficients have heavier tails and more sharply peaked modes at
zero compared with that of a Gaussian PDF. Here, using
Lagrange-multiplier test, we demonstrate that heteroscedasticity
exists in the contourlet coefficients, i.e., their conditional
variance is non-constant. This property has not been mentioned
before and the proposed models for these coefficients suppose them
to be identically distributed \cite{item888,item84}. To overcome
the inefficiency of the previously proposed models in capturing
the heteroscedasticity and the dependency of the contourlet
coefficients, we propose using 2D-GARCH model. In the following
parts, we review 2D-GARCH model and study whether this model
provides a flexible and appropriate tool for the coefficients
within the framework of multiscale and directional contourlet
analysis of the images. It should be mentioned that in this paper,
we use 9-7 biorthogonal filters for multiscale decomposition, and
PKVA ladder filters \cite{itemc6} for multi directional
decomposition \cite{item1c}. 
\subsection{2D-GARCH Model}
Generalized autoregressive conditional heteroscedasticity (GARCH)
model has been proposed by Bollerslev in \cite{item16}. GARCH
processes are a class of zero mean, serially uncorrelated, but not
serially independent processes with non-constant variances
conditioned on the past \cite{item16}. 2D-GARCH processes are the
extension of GARCH in two dimensions \cite{item12p}. Suppose
$f_{ij}$ represents a two dimensional stochastic process that
follows 2D-GARCH($p_1,p_2$ ,$q_1,q_2$), where ($p_1,p_2$
,$q_1,q_2$) denotes the order of the model. We have
\begin{eqnarray}
\label{eq04}
f_{ij}&=& \sqrt{h_{ij}} \varepsilon_{ij}\\
\label{eq05} h_{ij}  &=& \alpha _0  + \sum_{k\ell \in \Omega _1 }
{\alpha _{k\ell} f_{i - k,j - \ell}^2 }  + \sum_{k\ell \in \Omega
_2 } {\beta _{k\ell} h_{i - k,j - \ell}^{} },
\end{eqnarray}
where $\varepsilon_{ij}$ represents an i.i.d two-dimensional stochastic process with standard normal distribution ($\varepsilon_{ij}$$\sim$$\mathcal{N}(0,1)$) and
\begin{eqnarray*}
\Omega _1  &=& \{ k\ell\left| {0 \leq  k \leq q_1 ,0
\leq \ell \leq q_2 } \right.,(k\ell) \ne (0,0)\}
\\
\Omega _2 &=& \{ k\ell\left| {0 \leq k \leq p_1 ,0
\leq \ell \leq p_2 } \right.,(k\ell) \ne (0,0)\}.
\end{eqnarray*}
The conditional variance of $f_{ij}$ is $h_{ij}$ and the
conditional distribution of $f_{ij}$ can be formulated as:
\begin{equation}
\label{e3} p(f_{ij} \left| {\psi _{ij} } \right.) =
\frac{1}{{\sqrt {2\pi h_{ij} } }}\exp (\frac{{ - f_{ij} ^2 }}{{2h_{ij} }}),
\end{equation}
where $\psi_{ij}$ is the information set defined as \[
\psi _{ij}  = \{ \{ f_{i - k,j - \ell} \} _{k,\ell \in \Omega _1
} ,\{ h_{i - k,j - \ell} \} _{k,\ell \in \Omega _2 } \}.
\]The model
parameters are $\Gamma$  = $\{ \{ \alpha _0$, $\alpha _{01}$ ,
$\cdots,\alpha _{q_1 q_2 } ,\beta _{01} ,\cdots,\beta _{p_1 p_2 }
\} \} $ that should be estimated. We use maximum likelihood
estimation and the likelihood function can be formulated as
\begin{equation}
LF(\Gamma ) = \prod\limits_{ij \in \Phi } {p(f_{ij} \left|
{\psi _{ij} )} \right.},
\end{equation}
where $\Phi  = \{ ij| {1 \leq  i \leq  M,1 \leq j \leq
N\} }$ is the sample space of size $M\times N$.
\begin{table}
\begin{center}
\caption{\label{table1}  Results of  Engle and two dimensional
\cite{item24} hypothesis tests for the presence of 2D-GARCH effect
in the eight contourlet subbands of the finest scale for the {\it
Peppers} image.}
\begin{tabular}{c|c|c|c|c}
 \hline   &  scan &  H & pValue  & \ \ GARCHstat \ \ \\
 \hline
 \hline
 & vertical&1&0&941.8335\\
  \cline{2-5}
subband  &horizontal&1&0&3.9431e+003\\
 \cline{2-5}
 1&diagonal&1&0&5.1349e+003\\
  \cline{2-5}
  \cline{2-5}
  &two dimensional&1&0&1.0166e+004\\
\hline \hline
& vertical&1&0&2.6268e+004\\
  \cline{2-5}
subband  &horizontal&1&0&2.3267e+003\\
 \cline{2-5}
 2&diagonal&1&0&633.5300\\
  \cline{2-5}
  \cline{2-5}
  &two dimensional&1&0&2.1819e+004\\
\hline \hline
& vertical&1&0&7.7814e+003\\
  \cline{2-5}
subband  &horizontal&1&0&2.4306e+003\\
 \cline{2-5}
 3&diagonal&1&0&3.3464e+003\\
  \cline{2-5}
  \cline{2-5}
  &two dimensional&1&0&7.1432e+003\\
\hline \hline
& vertical&1&0&553.8294\\
  \cline{2-5}
subband  &horizontal&1&0&1.7055e+003\\
 \cline{2-5}
 4&diagonal&1&0&1.3882e+003\\
  \cline{2-5}
  \cline{2-5}
  &two dimensional&1&0&6.8305e+003\\
\hline \hline
& vertical&1&0&1.5818e+003\\
  \cline{2-5}
subband  &horizontal&1&0&71.1390\\
 \cline{2-5}
 5&diagonal&1&0&2.9839e+003\\
  \cline{2-5}
  \cline{2-5}
  &two dimensional&1&0&1.2155e+004\\
\hline \hline
& vertical&1&0&3.1624e+003\\
  \cline{2-5}
subband  &horizontal&1&0&1.1515e+004\\
 \cline{2-5}
 6&diagonal&1&0&2.8019e+003\\
  \cline{2-5}
  \cline{2-5}
  &two dimensional&1&0&6.9759e+003\\
\hline \hline
& vertical&1&0&6.2005e+003\\
  \cline{2-5}
subband  &horizontal&1&0&2.6225e+004\\
 \cline{2-5}
 7&diagonal&1&0&1.7621e+003\\
  \cline{2-5}
  \cline{2-5}
  &two dimensional&1&0&2.5357e+004\\
\hline \hline
& vertical&1&0&2.5334e+003\\
  \cline{2-5}
subband  &horizontal&1&0&1.4457e+003\\
 \cline{2-5}
 8&diagonal&1&0&5.9819e+003\\
  \cline{2-5}
  \cline{2-5}
  &two dimensional&1&0&8.1015e+003\\
\hline \hline
\end{tabular}
\end{center}
\end{table}

\subsection{2D-GARCH Modeling of the Contourlet Coefficients}
Here, we study the compatibility of 2D-GARCH model with the
contourlet coefficients.
 In this way, we carried out extensive simulations on a large number of images.
 But, due to the space limitation, we present some limited results. 2D-GARCH is a heteroscedastic model that allows the conditional variance to change over two dimensions with a special structure of the dependencies as described in section 3.1.
 To check the suitability of the 2D-GARCH model for the contourlet coefficients, we should examine the heteroscedasticity of these coefficients and
 their compatibility with the special type of dependency
provided by 2D-GARCH model. In this way, two LM tests have been proposed before \cite{item23,item24} and we use them:\\
1) LM test proposed by Engle in \cite{item23}: This test checks
the null hypothesis that no GARCH effects exist. It can be used
for one dimensional signals. Comparing the structure of
conditional variance in 1D-GARCH and 2D-GARCH models, we can
conclude that to test the two dimensional GARCH effect, we can
apply it for horizontal, vertical, and diagonal scans of the
contourlet subbands.\\2)  LM test  proposed
in \cite{item24,item24m} that examines the two-dimensional GARCH effect.\\
The results of applying these two hypothesis tests for the eight
directional subbands in the finest scale of the {\it Peppers}
image have been shown in table~\ref{table1}. For other test images, similar results have been
 obtained.  In this table ``H''
is a Boolean decision variable that ``1'' indicates acceptance of
the alternative hypothesis that GARCH effects exist, ``pValue'' is
the significance level at which this test rejects the null
hypothesis, and ``GARCHstat'' indicates GARCH test statistic. The
significance level is $0.05$ . This table demonstrates the
existence of two dimensional heteroscedasticity in the contourlet
coefficients that  can be  efficiently captured using 2D-GARCH
model.\\ To study the statistical significance of the reported
results, we perform the LM tests on the eight directional subbands
in the finest scale of 50 natural images. Table~\ref{table111}
reports the mean and standard deviation of the results. All of the
computed GARCH test statistics ``GARCHstat''s  are large. Since
the standard deviation of ``GARCHstat'' is also large, we report
the minimum and maximum value of ``GARCHstat'', too.   It is
obvious from this table that all of the tested subbands are
heteroscedastic. So, we should mention this property in modelling
the contourlet subbands .

\begin{table*}[!t]
\begin{changemargin}{-1cm}{-1cm}
\begin{center} {\small
\caption{\label{table111}  Results of  Engle and two dimensional
\cite{item24} hypothesis tests for the presence of 2D-GARCH effect
in the contourlet subbands of 50 natural images.}
\begin{tabular}{c|c|c|c|c|c|c|c|c}
 \hline     scan & \multicolumn{2} {c|} {H} & \multicolumn{2} {c|} {pValue}  &  \multicolumn{4} {c} {GARCHstat}  \\
  \cline{2-9}
 &mean&std&mean&std&mean&std&min&max\\
 \hline
 \hline
 vertical&1&0&0&0&4.83e+003&2.04e+003&589.23&1.00e+004\\
  \hline
horizontal&1&0&0&0&4.87e+003&2.23e+003&571.51&1.22e+004\\
 \hline
 diagonal&1&0&0&0&5.58e+003&2.28e+003&2.43e+003&1.42e+004\\
  \hline
  \hline
 two dimensional&1&0&0&0&1.11e+004&2.07e+003&7.29e+003&1.74e+004\\
\hline \hline
\end{tabular}}
\end{center}
\end{changemargin}
\end{table*}

\begin{figure*}[!t]
\begin{center}
\includegraphics[scale=.45]{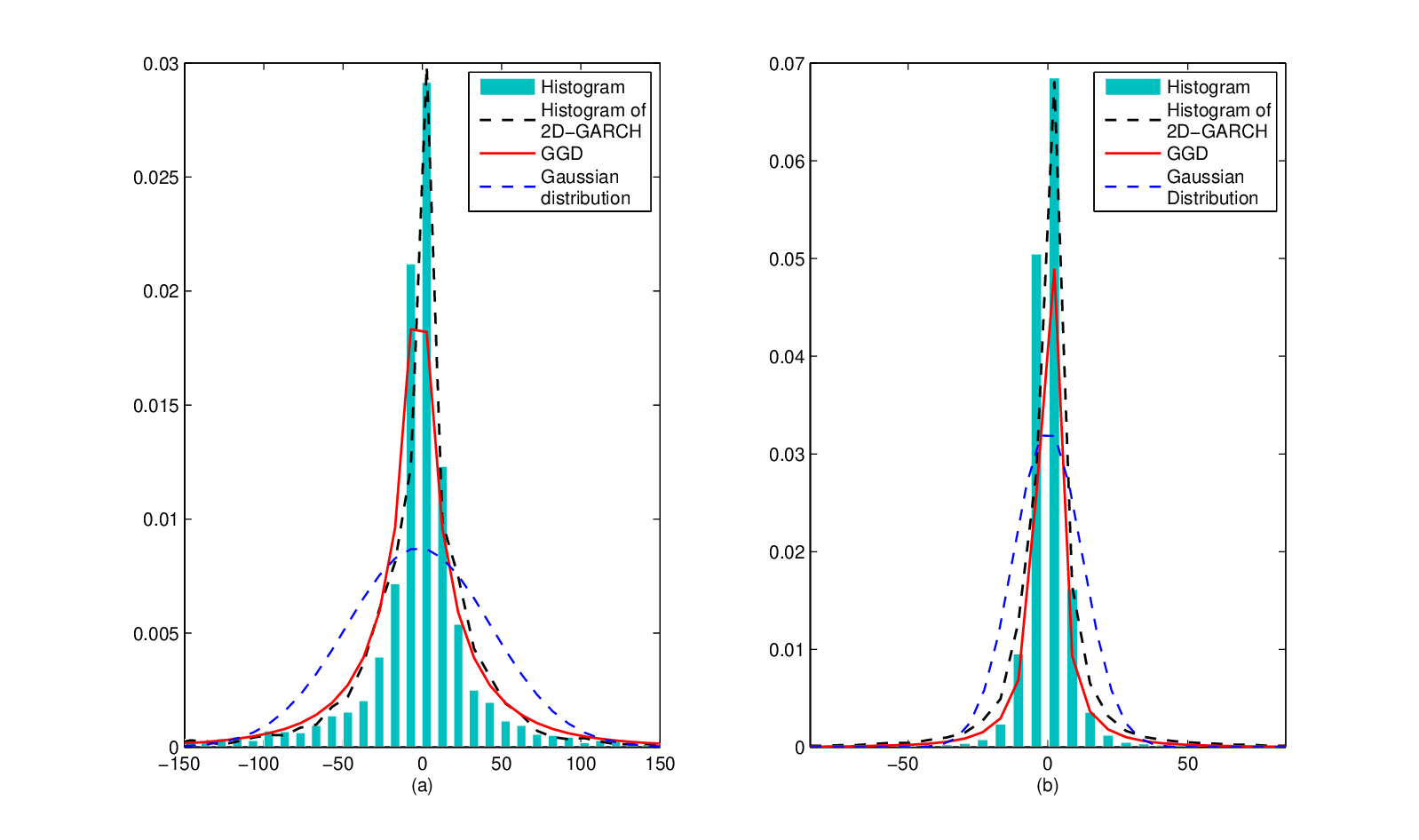}
\end{center}
\caption{\label{fig3} Histograms of the contourlet coefficients
for {\it Peppers} image: (a)  directional subband in the first
pyramidal level (b) directional subband in the third pyramidal
level}
\end{figure*}
Also, we check the compatibility between the histograms of contourlet coefficients and the 2D-GARCH model. Fig.~\ref{fig3}  presents the histograms of two contourlet
subbands of {\it Peppers} image and the histograms of the best
fitted 2D-GARCH model. This figure also shows the best fitted
Gaussian and Generalized Gaussian (GG) distributions.
It is clear from this figure that 2D-GARCH model provides a better fit to the data. We have obtained similar results for other test images. \\

\section{WATERMARK EMBEDDING}
To embed the watermark, we use an additive spread spectrum
 scheme in the contourlet domain. First, we apply the contourlet transform to the original image.
To increase the robustness of the watermark, we insert the
watermark in the most significant direction of the image. In this
way, we compute the energy of each directional subband in the
finest scale and select the subband with the highest energy to
embed the watermark. Let ${\bf{f}}=\{f_{ij}|i=1,...,M,j=1,...,N\}$
denotes the selected subband of the original image. We use bold
type to denote two dimensional vectors. The rule for additive
embedding of the watermark sequence in this subband is
\begin{equation}
\label{n1} g_{ij}=f_{ij}+\gamma s_{ij}=f_{ij}+w_{ij},
\end{equation}
where ${\bf{s}}=\{s_{ij}|i=1,...,M,j=1,...,N\}$ denotes the
watermark sequence used for marking the contourlet subband,
${\bf{g}}=\{g_{ij}|i=1,...,M,j=1,...,N\}$ denotes the watermarked
contourlet subband, and $\gamma$ is the embedding power.
${\bf{s}}$ is a bipolar watermark taking the values $-1$ and $1$
with the same probability. It is obtained by using a pseudorandom
sequence (PRS) generator with an initial state depends on the
value of a secret key. The final watermark signal $w_{ij}=\gamma
s_{ij}$  is  generated by the multiplication of the pseudorandom
sequence $s_{ij}$ and the embedding power $\gamma$ \cite{item5}.
$\gamma$  controls the watermark to document ratio (WDR).
\section{WATRMARK DETECTION BASED ON 2D-GARCH MODEL}
For copyright protection, the detector needs to verify the existence of a known watermark
 in a given image \cite{item5,item888}. Using  (\ref{n1}), the additive watermark detection for copyright protection
in the contourlet domain can be mathematically formulated as the
binary hypothesis test:
\begin{eqnarray}
\label{ee1}\mathcal{H}_0 &:& g_{ij}=f_{ij}\\
\label{ee2}\mathcal{H}_1 &:& g_{ij}=f_{ij}+ w_{ij}
\end{eqnarray}
in which, we verify the existence of $w_{ij}$  in the contourlet
coefficients of an image. Here, $\mathcal{H}_0$ and
$\mathcal{H}_1$ denote the null and alternative hypotheses,
respectively. In this work, we use a Bayesian log-likelihood ratio
test (LLRT) to detect the watermark. It should be mentioned that a
detector based on LLRT maximizes the probability of detection
(deciding $\mathcal{H}_1$ when $\mathcal{H}_1$ is true) for a
fixed probability of false-alarm (deciding $\mathcal{H}_1$ when
$\mathcal{H}_0$ is true). The Bayesian LLRT is given by
\begin{equation}
\label{e1} \log\{\Lambda({\bf{g}})\}=\log\{\frac{p({\bf{g}}|\mathcal{H}_1)}{p({\bf{g}}|\mathcal{H}_0)}\} {\begin{array}{c}1\\[-.1cm] > \\[-.2cm] < \\[-.1cm] 0 \end{array}} log\{ k\frac{P(\mathcal{H}_0)}{P(\mathcal{H}_1)}\}=T
\end{equation}
where $p({\bf{g}}|\mathcal{H}_0)$ and $p({\bf{g}}|\mathcal{H}_1)$
are the pdfs of ${\bf{g}}$ under the conditions $\mathcal{H}_0$
and $\mathcal{H}_1$, respectively, $k$ is a constant denotes the
ratio of false alarm cost to the miss-detection (deciding
$\mathcal{H}_0$ when $\mathcal{H}_1$ is true) cost.
$P(\mathcal{H}_0)$ and $P(\mathcal{H}_1)$ are the probabilities of
null and alternative hypotheses, respectively. To minimize the
probability of miss-detection for a bounded false alarm
probability,
the threshold $T$ is computed using Neyman–Pearson criteria.\\
To design an efficient detector using (\ref{e1}), the correct
choice of priors for the contourlet coefficients is certainly a
very important factor. As studied in section 3.2, there is a good
compatibility between 2D-GARCH model and the contourlet
coefficients. So, we design a watermark detector based on this
model.  Assume that the contourlet coefficients of the original
image ${\bf{f}}$ follow 2D-GARCH model. Using (\ref{e3}),
(\ref{ee1}), and (\ref{ee2}), the log-likelihood ratio given in
(\ref{e1}) can be written as:
\begin{eqnarray}
\label{e2} \log\{\Lambda({\bf{g}})\}&=&\log\frac{\prod\limits_{ij
\in \Phi } {p(g_{ij}-w_{ij} \left| {\psi _{ij} )}
\right.}}{\prod\limits_{ij \in \Phi } {p(g_{ij} \left|
{\psi _{ij} )} \right.}}\\
\label{e4}&=&\log\frac{\prod\limits_{ij \in \Phi }
{\frac{1}{{\sqrt {2\pi h_{ij} } }}\exp (\frac{{ - (g_{ij}-w_{ij})
^2 }}{{2h_{ij} }})}}{\prod\limits_{ij \in \Phi } {\frac{1}{{\sqrt
{2\pi h_{ij} } }}\exp (\frac{{ - g_{ij} ^2 }}{{2h_{ij} }})}}
\end{eqnarray}
where $\psi _{ij}$ and $\Phi$ are as defined in section 3.1.
Substituting $h_{ij}$ from (\ref{eq05}) in (\ref{e4}), the
log-likelihood ratio can be formulated  as (\ref{e5}).
\begin{changemargin}{-3.9cm}{3cm}

{\footnotesize
\begin{eqnarray}
 \nonumber  &\log&\{\Lambda({\bf{g}})\}=\log\frac{\prod\limits_{ij \in \Phi } {\frac{1}{{\sqrt {2\pi \alpha _0  + \sum_{k\ell \in \Omega _1 }
{\alpha _{k\ell} (g_{i - k,j - \ell}-w_{i - k,j - \ell})^2 }  +
\sum_{k\ell \in \Omega _2 } {\beta _{k\ell} h_{i - k,j - \ell}^{}
} } }}\exp (\frac{{ - (g_{ij}-w_{ij}) ^2 }}{{2\alpha _0  +
\sum_{k\ell \in \Omega _1 } {\alpha _{k\ell} (g_{i - k,j -
\ell}-w_{i - k,j - \ell})^2 }  + \sum_{k\ell \in \Omega _2 }
{\beta _{k\ell} h_{i - k,j - \ell}^{} } }})}}{\prod\limits_{ij \in
\Phi } {\frac{1}{{\sqrt {2\pi \alpha _0  + \sum_{k\ell \in \Omega
_1 } {\alpha _{k\ell} g_{i - k,j - \ell}^2 }  + \sum_{k\ell \in
\Omega _2 } {\beta _{k\ell} h_{i - k,j - \ell}^{} } } }}\exp
(\frac{{ - g_{ij} ^2 }}{{2\alpha _0  + \sum_{k\ell \in \Omega _1 }
{\alpha _{k\ell} g_{i - k,j - \ell}^2 }  + \sum_{k\ell \in \Omega
_2 } {\beta _{k\ell} h_{i - k,j - \ell}^{} } }})}} \\
\nonumber&=& \sum\limits_{ij \in \Phi } -\log \sqrt{{\sum_{k\ell
\in \Omega _1 } {\alpha _{k\ell} (g_{i - k,j - \ell}-w_{i - k,j -
\ell})^2 }
 + \sum_{k\ell \in \Omega
_2 } {\beta _{k\ell} h_{i - k,j - \ell}^{} }}}+ \log \sqrt{{\sum_{k\ell \in \Omega _1 }
{\alpha _{k\ell} g_{i - k,j - \ell}^2 }
 + \sum_{k\ell \in \Omega
_2 } {\beta _{k\ell} h_{i - k,j - \ell}^{} }}}  \\\nonumber
&+&\sum\limits_{ij \in \Phi }\left[\frac{{  g_{ij} ^2 }}{{2\alpha
_0 + \sum_{k\ell \in \Omega _1 } {\alpha _{k\ell} g_{i - k,j -
\ell}^2 }  + \sum_{k\ell \in \Omega _2 } {\beta _{k\ell} h_{i -
k,j - \ell}^{} } }}+\frac{{ - (g_{ij}-w_{ij}) ^2 }}{{2\alpha _0 +
\sum_{k\ell \in \Omega _1 } {\alpha _{k\ell} (g_{i - k,j -
\ell}-w_{i - k,j - \ell})^2 }  + \sum_{k\ell \in \Omega
_2 } {\beta _{k\ell} h_{i - k,j - \ell}^{} } }} \right]  \\
\nonumber&=& \sum\limits_{ij \in \Phi } \log
\sqrt{\frac{\sum_{k\ell \in \Omega _1 } {\alpha _{k\ell} g_{i -
k,j - \ell}^2 }  + \sum_{k\ell \in \Omega _2 } {\beta _{k\ell}
h_{i - k,j - \ell}^{} }}{\sum_{k\ell \in \Omega _1 } {\alpha
_{k\ell} (g_{i - k,j - \ell}-w_{i - k,j - \ell})^2 }  +
\sum_{k\ell \in \Omega _2 } {\beta _{k\ell} h_{i - k,j - \ell}^{}
}}}+\sum\limits_{ij \in \Phi }[ \frac{{  g_{ij} ^2 }}{{2\alpha _0
+ \sum_{k\ell \in \Omega _1 } {\alpha _{k\ell} g_{i - k,j -
\ell}^2 }  + \sum_{k\ell \in \Omega _2 } {\beta _{k\ell} h_{i -
k,j - \ell}^{} } }}  \\ \label{e5}  &+&\frac{{ - (g_{ij}-w_{ij})
^2 }}{{2\alpha _0  + \sum_{k\ell \in \Omega _1 } {\alpha _{k\ell}
(g_{i - k,j - \ell}-w_{i - k,j - \ell})^2 }  + \sum_{k\ell \in
\Omega _2 } {\beta _{k\ell} h_{i - k,j - \ell}^{} } }} ]
\end{eqnarray}
}
\end{changemargin}

\begin{figure*}[!t]
\begin{center}
\includegraphics[scale=0.45]{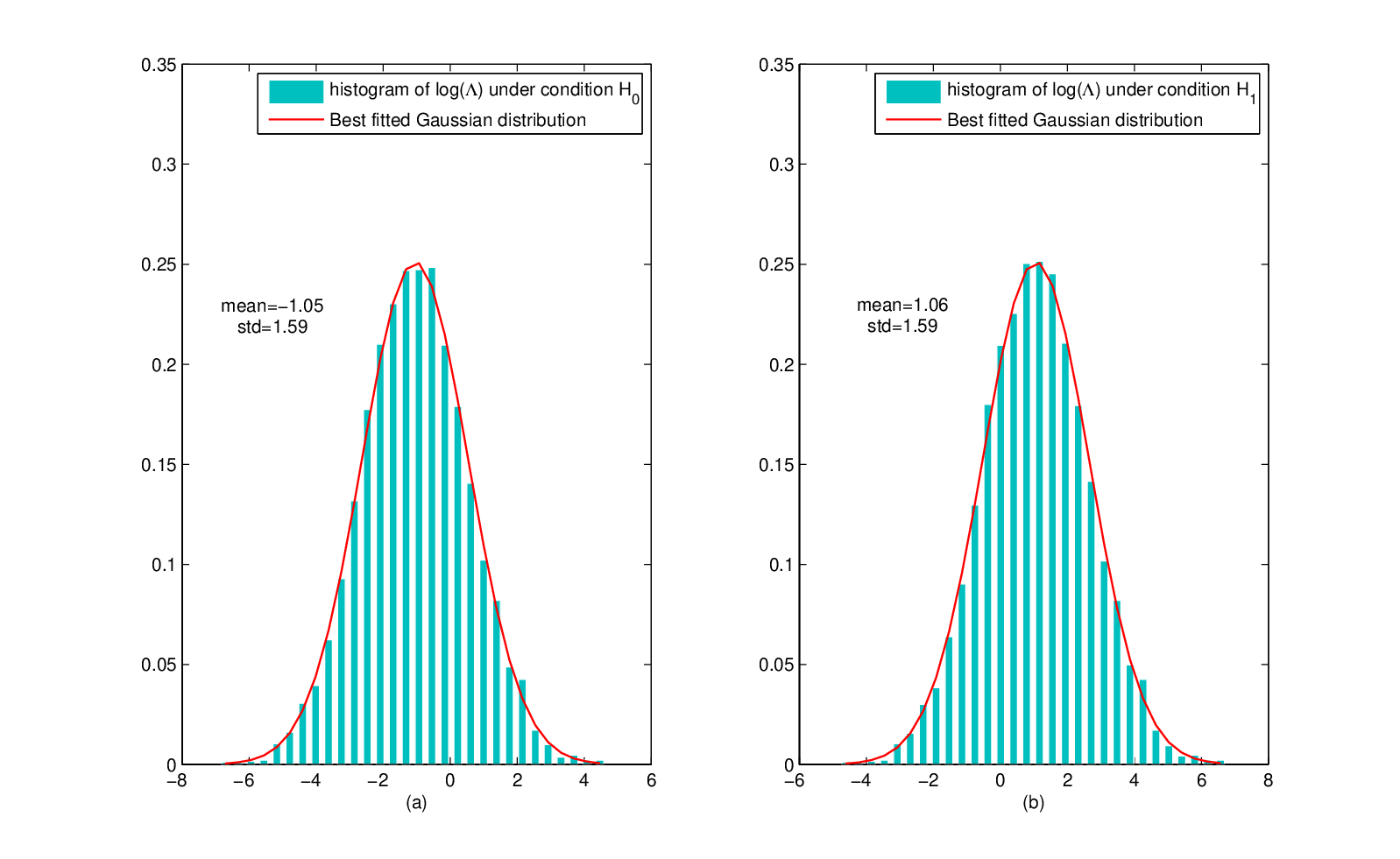}
\caption{\label{fig4} Histograms of log-likelihood ratio $\log
\Lambda$  under (a) $\mathcal{H}_0$ and (b) $\mathcal{H}_1$
conditions. Vertical bars show the normalized histogram , Gaussian
distribution depicted in red dashed line. (Peppers image,
directional subband in the finest scale with the highest energy) }
\end{center}
\end{figure*}

\subsection{Analysis of the Detector Performance}
To analyze the performance of our proposed detector
 in terms of its probability of false alarm $P_F$ and its probability
of detection $P_D$, first, we should investigate the distribution
of log-likelihood ratio $\log \Lambda$ in (\ref{e5}). In the
following, by using normalized histograms, Kolmogrov-Smirnov test
and the estimated kurtosis, we demonstrate that $\log \Lambda$
could be well approximated by a Gaussian distribution. \\
We obtain the histogram of log-likelihood ratio under the
conditions $\mathcal{H}_0$ and $\mathcal{H}_1$ by repeating the
embedding step 1000 times. Each time starts with a
uniquely-defined key to randomly generate the watermark sequence.
In the rest of this paper, we use a similar way to obtain the
experimental results. We focus on the contourlet subband in which
the watermark is embedded, i.e, the directional subband in the
finest scale with the highest energy. The histograms of
log-likelihood ratio for the most energetic subband in the finest
scale of the {\it Peppers} image under the conditions
$\mathcal{H}_0$ and $\mathcal{H}_1$ have been shown in
Fig.~\ref{fig4}.a and Fig.~\ref{fig4}.b, respectively (WDR= -50
dB). This figure also represents the best fitted Gaussian
distribution. We can see that the Gaussian distribution has been
well fitted to the histograms. Also, it is clear from
Fig.~\ref{fig4} that   the mean of $\log \Lambda$ under the
conditions $\mathcal{H}_0$ and $\mathcal{H}_1$ have approximately
the same amplitude and opposite sign, and the variances under
these conditions  are approximately equal. It should be mentioned
that the modeling results of different
images are similar. \\
\begin{table}
\begin{center}
\caption{\label{table2}  Results of the KS test
 for $\log \Lambda$ of Peppers image
 (directional subband in the finest scale with the highest energy) under the conditions $\mathcal{H}_0$ and $\mathcal{H}_1$.}
\begin{tabular}{c|c|c|c|c|c|c|c}
\hline
 \hline
  \multicolumn{2} {c|} {WDR  (dB)}  &  -50  &  -52 &-54  & -56   &  -58&-60 \\
    \hline
 \hline
 &$H$ &0&0&0&0&0&0\\
  \cline{2-8}
$\mathcal{H}_0$&$KSD$&0.0178&0.0166&0.0176&0.0169&0.0157&0.0150\\
 \hline
 \hline
  &$H$ &0&0&0&0&0&0\\
 \cline{2-8}
$\mathcal{H}_1$&$KSD$&0.0168&0.0172&0.0167&0.0168&0.0151&0.0152\\
 \hline
\hline
\end{tabular}
\end{center}
\end{table}
Now, we use the Kolmogrov-Smirnov (KS) test to quantify the
results. This test evaluates the compatibility between the
distribution of a sample data $p(x)$ and a given PDF $p_1(x)$. KS
method is a binary hypothesis test. The null hypothesis ($H_0$)
denotes that the distribution $p(x)$ is same as $p_1(x)$. To
employ this test, first, the KS distance ($KSD$)
 should be computed:
\begin{equation}
\label{eq7265}KSD = \max_{-\infty<x<\infty} |P(x) - P_1(x)|.
\end{equation}
where $P(x)$ and $P_1(x)$ denote the corresponding cumulative
distribution functions. Then, KS distance is compared with a
threshold to decide between two hypotheses ($H_0$ and $H_1$). This
threshold is determined based on the significance level of the
test. We perform KS test to examine the compatibility between the
density of  $\log \Lambda$ ($p(x)$) with the Gaussian distribution
($p_1(x)$).
 Table~\ref{table2} reports the accepted hypothesis ``$H$" and $KSD$ in different WDRs for the {\it Peppers} image. ``$H$" takes the values ``0" and ``1" to indicate $H_0$ and $H_1$, respectively. These results verify that $\log \Lambda$ can
be well approximated with the Gaussian distribution.\\ Also, the
sample kurtosis of  $\log \Lambda$ (fourth moment divided by the
square of the second moment)  in different WDRs have been reported
in table~\ref{table3}. All of the estimated kurtoses are close to
the value of three, which is expected for a Gaussian distribution.
\\ Finally, to demonstrate the statistical significance of the
reported results, we perform the KS test on  $\log \Lambda$ of
contourlet coefficients for 50 natural images (the subbands in the
finest scale with the highest energy) and also compute the
corresponding kurtoses.  The mean and standard deviation of the
results have been reported in table~\ref{table22} and
table~\ref{table33}. It is evident from these tables that  for all
of the tested subbands, $\log \Lambda$ can be efficiently
approximated by the Gaussian distribution.
\begin{table}
\begin{center}
\caption{\label{table3}  kurtosis
 for $\log \Lambda$ of Peppers image (directional subband in the finest scale with the highest energy) under the conditions $\mathcal{H}_0$ and $\mathcal{H}_1$. }
\begin{tabular}{c|c|c|c|c|c|c}
\hline
   \hline
   {WDR  (dB)}  &  -50  &  -52 &-54  & -56   &  -58&-60 \\
    \hline
 \hline
$\mathcal{H}_0$ &2.9346
&3.0898&2.9808&2.9213&2.9451&2.8006\\
 \hline
 \hline
$\mathcal{H}_1$ &2.9370&3.0902&2.9810&2.9197&2.9450&2.8012\\
 \hline
\hline
\end{tabular}
\end{center}
\end{table}

\begin{table*}[!t]
\begin{center}
\caption{\label{table22} mean and standard deviation of the KS
test results  for $\log \Lambda$ of 50 natural images (directional
subband in the finest scale with the highest energy) under the
conditions $\mathcal{H}_0$ and $\mathcal{H}_1$.}
\begin{tabular}{c|c|c|c|c|c|c|c}
\hline \hline \multicolumn{8}{c} {mean}\\ \hline \hline
  \multicolumn{2} {c|} {WDR  (dB)}  & { -50}  &  {-52} &{-54 } & {-56}   &  {-58}& {-60} \\
       \hline
 \hline
 &$H$ &0&0&0&0&0&0\\
 \cline{2-8}
$\mathcal{H}_0$&$KSD$&0.0203&0.0212& 0.0196&0.0201&  0.0199&0.0214 \\
 \hline
 \hline
  &$H$ &0&0&0&0&0&0\\
  \cline{2-8}
 $\mathcal{H}_1$&$KSD$&0.0202& 0.0207& 0.0197&0.0202& 0.0198 &0.0212\\
 \hline
\hline \multicolumn{8}{c} {std}\\ \hline \hline
  \multicolumn{2} {c|} {WDR  (dB)}  & { -50}  &  {-52} &{-54 } & {-56}   &  {-58}& {-60} \\
       \hline
 \hline
 &$H$ &0&0&0&0&0&0\\
 \cline{2-8}
$\mathcal{H}_0$&$KSD$&0.0060& 0.0056&0.0048&0.0051&  0.0050&0.0062\\
 \hline
 \hline
  &$H$ &0&0&0&0&0&0\\
  \cline{2-8}
 $\mathcal{H}_1$&$KSD$&0.0053&0.0051& 0.0051& 0.0056 & 0.0051&0.0058\\
 \hline
\hline
\end{tabular}
\end{center}
\end{table*}

\begin{figure*}[!t]
\begin{center}
\includegraphics[scale=.52]{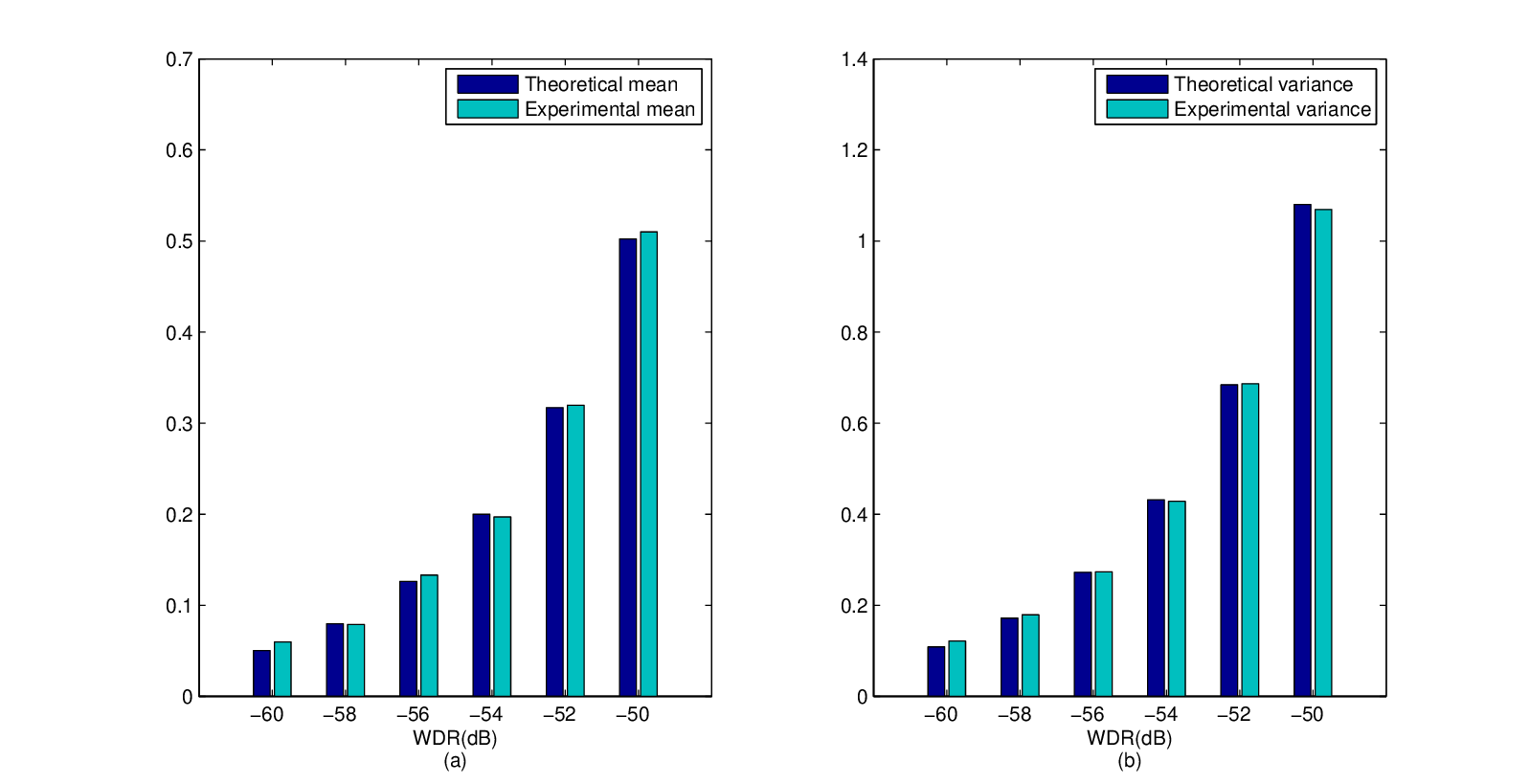}
\end{center}
\caption{\label{fig5} (a) Empirical and theoretical means of test
statistic, (b) Empirical and theoretical variances of test
statistic. (Peppers image, directional subband in the finest scale
with the highest energy) } \vspace*{4pt}
\end{figure*}

\begin{figure}
\begin{center}
\includegraphics[scale=.35]{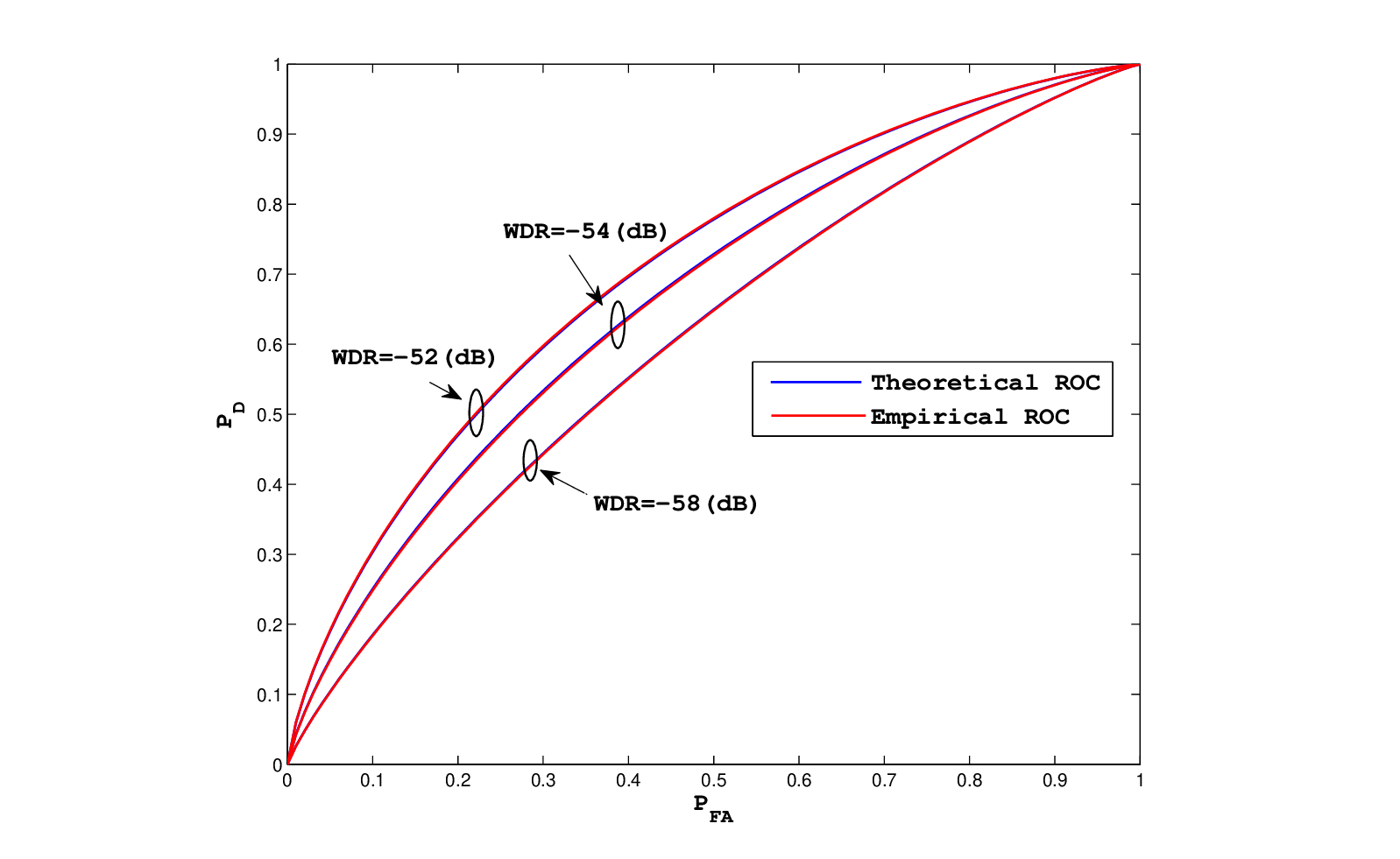}
\end{center}
\caption{\label{fig6} Empirical and theoretical ROCs (Peppers
image, directional subband in the finest scale with the highest
energy) } \vspace*{4pt}
\end{figure}

\begin{table*}[!t]
\begin{center}
\caption{\label{table33}   mean and standard deviation of kurtosis
 for $\log \Lambda$ of 50 natural image (directional subband in the finest scale with the highest energy) under the conditions $\mathcal{H}_0$ and $\mathcal{H}_1$. }
\begin{tabular}{c|c|c|c|c|c|c}
\hline \hline \multicolumn{7}{c} {mean}\\ \hline \hline
  {WDR  (dB)}  & { -50}  &  {-52} &{-54 } & {-56}   &  {-58}& {-60} \\
       \hline
 \hline

$\mathcal{H}_0$&2.9146 & 2.8911&2.9485&2.9305
& 2.9326& 2.9474\\
 \hline
 \hline
  $\mathcal{H}_1$&2.9181&2.8900&2.9506&2.9357
& 2.9325&2.9608\\
 \hline
\hline \multicolumn{7}{c} {std}\\ \hline \hline
   {WDR  (dB)}  & { -50}  &  {-52} &{-54 } & {-56}   &  {-58}& {-60} \\
       \hline
 \hline

$\mathcal{H}_0$&0.2301&
 0.1672& 0.1614
&0.2143& 0.2126& 0.1710\\
 \hline
 \hline
  $\mathcal{H}_1$&0.2248& 0.1664&0.1722
&0.2229& 0.2129& 0.1674\\
 \hline
\hline
\end{tabular}
\end{center}
\end{table*}
So, we assume that $\log \Lambda$ follows the Gaussian
distribution. Based on this assumption, the probability of false
alarm and the probability of detection can be computed as
\cite{item5}:
\begin{equation}
\label{e13} P_F=Q\left(\frac{T-\mu_0}{\sigma_0}\right) \quad
P_D=Q\left(\frac{T-\mu_1}{\sigma_1}\right)
\end{equation}
where $Q(x)$ is defined as
$Q(x)=\frac{1}{\sqrt{2\pi}}\int_{x}^{\infty} \exp(-\frac{u^2}{2})
du$ and $T$ is the threshold in (\ref{e1}).
$\mu_0,\sigma_0,\mu_1,\sigma_1$ are the mean and standard
deviation of $\log \Lambda$ conditioned on the two hypotheses,
$\mathcal{H}_0$ and $\mathcal{H}_1$.  From (\ref{e13}), we have
\begin{equation}
\label{eee1} T= \sigma_0 Q^{-1}(P_F)+\mu_0
\end{equation}
Using (\ref{e13}) and (\ref{eee1}), the receiver operating
characteristic (ROC) of the watermark detector that presents the
function relation between the probabilities of detection ($P_D$)
and false alarm ($P_F$ ) can be formulated as
\begin{eqnarray}
\label{e14} P_D=Q\left(\frac{1}{\sigma_1}\left[\sigma_0
Q^{-1}(P_F)+\mu_0-\mu_1\right]\right).
\end{eqnarray}
Therefore, to analyze the performance of the proposed detector,
$\mu_0,\sigma_0,\mu_1,$ and $\sigma_1$ should be estimated. In Appendix A, we compute them. \\
We should notice that the watermark signal is very small in comparison with the original contourlet coefficients. Therefore, similar to some previously proposed watermark detectors such as \cite{item5,item10}, we suppose that the insertion of the watermark doesn't change the parameters of the statistical model significantly.
 We estimate the parameters of 2D-GARCH model using the received image, so, the proposed watermark detector is blind.\\
Here, we examine the validity of the theoretical mean and variance
of $\log \Lambda$ under $\mathcal{H}_0$ and $\mathcal{H}_1$
computed in  (\ref{e66})-(\ref{e111}) by using the mean and
variance of the experimental results of the test statistic. Since
we have $\mu_0=-\mu_1,  \quad \sigma_0^2=\sigma_1^2$, reporting
the results under one of the hypotheses is sufficient. We
performed Monte Carlo experiments using the 2D-GARCH detector and
compute the mean and the variance of the experimental results. The
experimental and the theoretical means and variances as a function
of WDR have been plotted in Fig.~\ref{fig5}.a and
fig.~\ref{fig5}.b.  Also, Fig.~\ref{fig6} shows the experimental
and the theoretical ROCs for three different WDRs. It is clear
from Fig.~\ref{fig5} and Fig.~\ref{fig6} that the empirical
measurements fit the theoretical estimates.
\section{EXPERIMENTAL RESULTS}
In this section, we study the performance of the proposed
watermark detector experimentally. For multiscale decomposition,
9-7 biorthogonal filters with two levels of pyramidal
decomposition are used. In the multi directional decomposition
stage, PKVA ladder filters are employed. The finest scale is
decomposed into eight directional subbands and the subband with
the highest energy is selected to embed the watermark. We use
2D-\GARCH(1,1,1,1).
To evaluate the efficiency of the proposed watermark detector, we
performed simulations on a large number of images. However, due to
the space limitations, we report the results of four $512\times
512$ grayscale representative images, namely, {\it Peppers, Living
room, Lake}, and {\it Pirate}. Also, we used 50 natural images
and report the averaged results. In the following, first, we
assess the performance of the contourlet domain 2D-GARCH detector
without any kind of attack. Then, since copyright protection needs robust watermark detector, we study the robustness of the
proposed detector under different kinds of attacks. Also, we
compare the proposed method with the other related detectors.
\subsection{Performance without attack}
\begin{figure*}[!t]
\begin{center}
\includegraphics[scale=0.55]{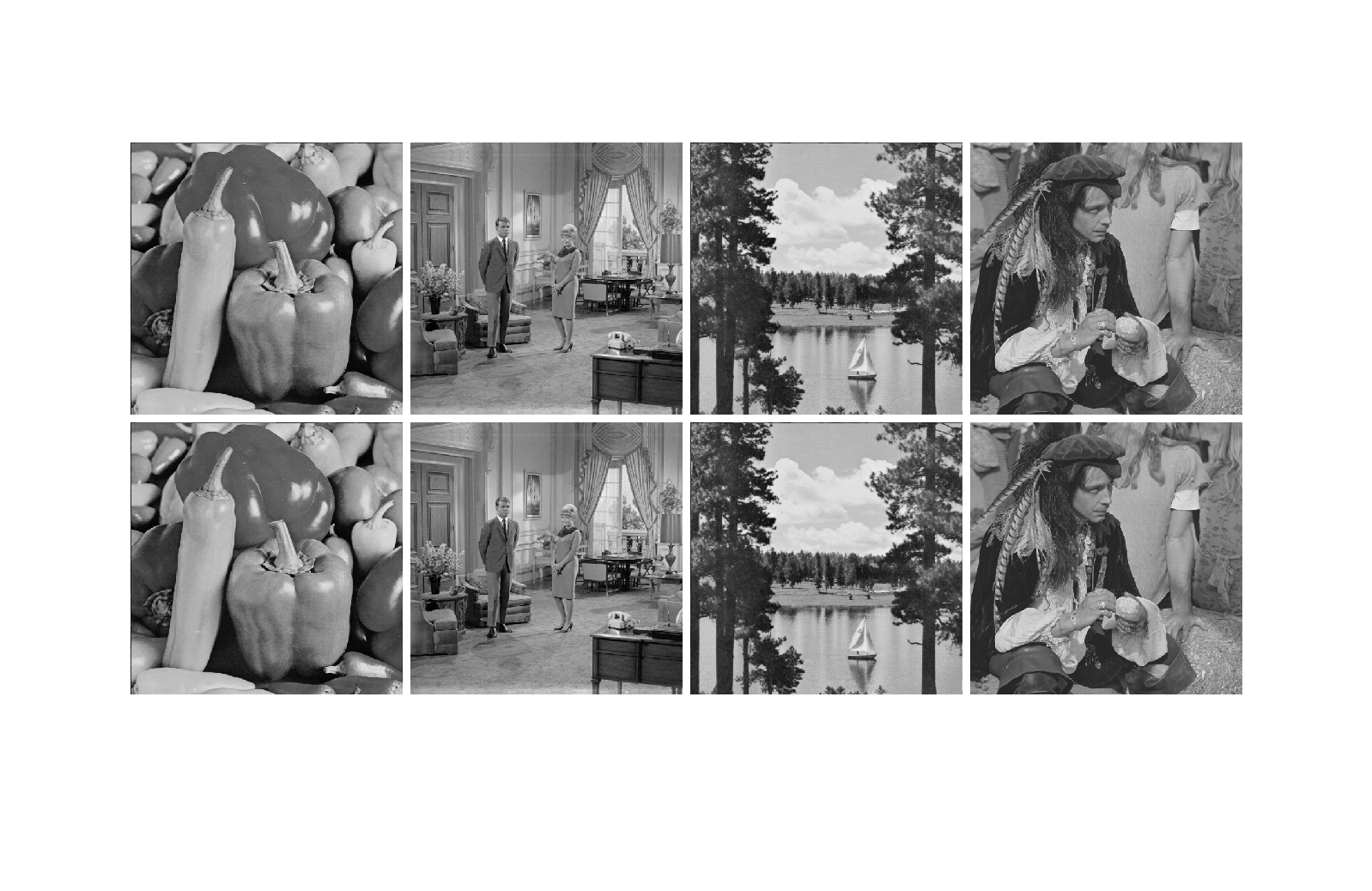}
\end{center}
\caption{\label{fig7} Test images: ``Peppers", ``Living
Room",`Lake`", and ``Pirate", up to down: Original images, and
watermarked images using the proposed scheme with WDR=-50 dB }
\end{figure*}

\begin{figure*}[!t]
\begin{center}
\includegraphics[scale=.55]{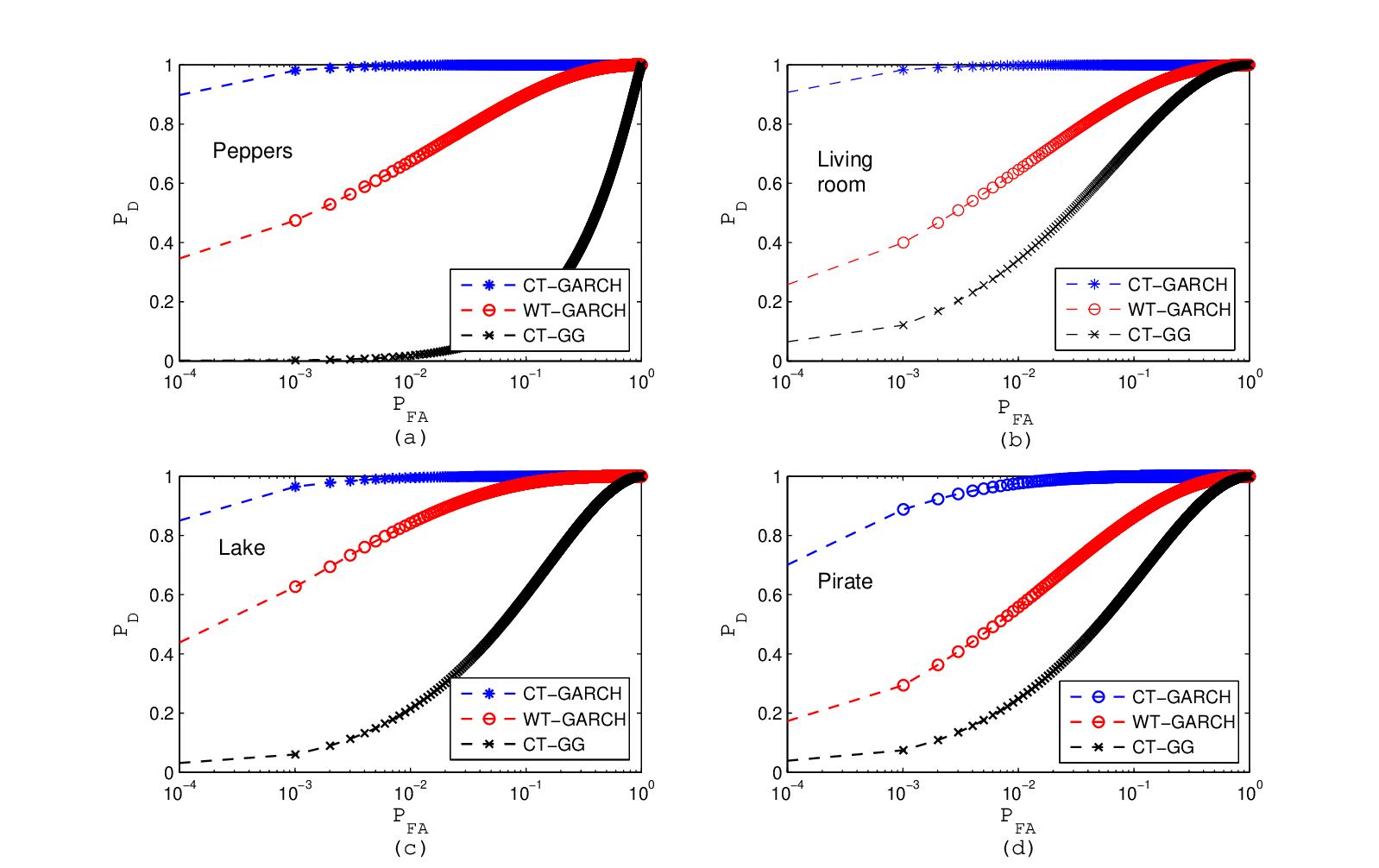}
\end{center}
\caption{\label{fig8} ROC of the three statistical detectors. Test
images are (a) Peppers, (b) Living Room, (c) Lake, and (d) Pirate.
(WDR=-50 dB) } \vspace*{4pt}
\end{figure*}

In fig.~\ref{fig7}, four test images and the watermarked version
of them with {\small$ WDR=-50 dB$} are represented. This figure
confirms that the proposed method satisfies the watermark
invisibility.\\
Now, we use ROC to study the performance of the proposed detector
and to compare it with the other related detectors.
Fig.~\ref{fig8} presents the ROCs of the contourlet domain
2D-GARCH based detector ({\small CT-GARCH}) for the four test
images. This figure also compares the ROCs of the proposed method
with two other detectors: 1) contourlet domain generalized
Gaussian based detector (CT-GG). It should be mentioned that using
generalized Gaussian distribution for the contourlet coefficients
has been proposed in some papers such as \cite{item84}. 2) wavelet
domain 2D-GARCH based detector ({\small WT-GARCH}). For the
wavelet domain 2D-GARCH based detector, 2D-\GARCH(1,1,1,1) is used
and the watermark is embedded in the detail subbands of the second
level decomposition. We use ``Daubechies" wavelet with four
vanishing moments.  This figure shows that the proposed detector
provides the highest probability of detection for any chosen value
of the false alarm.
\subsection{Performance under attacks}
In this section, we assess the performance of the contourlet
domain 2D-GARCH based detector under some standard attacks, i.e.,
JPEG compression, median and Gaussian filtering, scaling, and a
combinational attack. Also, we compare the performance of the
{\small CT-GARCH} detector with the {\small CT-GG} and {\small
WT-GARCH} detectors. To study the detection performance
quantitatively, the area under the ROC curve (AUROC) has been used
\cite{item10,item25}.\\ Table~\ref{table4} reports the AUROC
results of the {\small CT-GARCH},  {\small CT-GG}, and {\small
WT-GARCH} detectors under JPEG compression attack for WDR = -45 dB
and WDR = -50 dB. It is clear from this table that the {\small
CT-GARCH} detector outperforms the other
detectors.\\
\begin{table}
\begin{center}
\caption{\label{table4} auroc results under JPEG compression
(QF=60) attack }
\begin{tabular}{c |c |c |c }
\hline
 \hline
  \multicolumn{4} {c} {WDR = -50 dB} \\
 \hline
 \hline
     Image & {\small CT-GG }& {\small WT-GARCH }&  {\small CT-GARCH}   \\
     &   Detector&  Detector&  Detector  \\
\hline
  Peppers &0.7887&0.8512&0.9814\\
  \hline
  Living room&0.7527&0.9294&0.7527\\
  \hline
  Lake&0.8968&0.0.8393&0.9488 \\
  \hline
 Pirate &0.6818&0.8166&0.9145\\
   \hline
\hline
\multicolumn{4} {c} {WDR = -45 dB} \\
 \hline
 \hline
       Peppers &0.8634&0.9407&1.0000\\
  \hline
  Living room&0.9718&0.9765&1.0000\\
  \hline
  Lake&0.9916&0.9338&0.9964 \\
  \hline
 Pirate&0.8449&0.9143&0.9948\\
   \hline
\hline
\end{tabular}
\end{center}
\end{table}
The AUROC results of the {\small CT-GARCH},  {\small CT-GG}, and
{\small WT-GARCH} detectors under scaling attack have been
reported in table~\ref{table5} (WDR = -45 dB and WDR = -50 dB). We
can see the higher
performance of {\small CT-GARCH} detector under scaling attack. \\
The median filter is a nonlinear filter which produces a smoother
image. This filter might cause to fail in watermark detection and
a detector performance under this attack is demanding.
Table~\ref{table6} and table~\ref{table7} represent the AUROC
results of the detectors in WDR = -45 dB and WDR = -50 dB under
median and Gaussian filtering attacks, respectively  (with the
windows of size $5 \times 5$). For Gaussian filtering, the
standard deviation of the filter is set to $w/6$ where $w$ is the
size of window. Due to the very high efficiency of the proposed
method under Gaussian filtering attack in WDR = -45 dB and WDR =
-50 dB, we also report the results under this attack for very weak
watermark signal WDR = -60 dB. The reported results
demonstrate the efficiency of the {\small CT-GARCH} detector. \\
To investigate the performance of the detectors under
combinational attacks, table~\ref{table8} reports the AUROC
results under combination of the Gaussian filtering (window size =
$5 \times 5$) and additive white Gaussian noise (AWGN) attacks
(std=10). The high performance of the proposed method
under this combinational attack is obvious from this table.\\
Finally, to investigate the statistical significance of the
reported results, we perform the robustness tests on 50 natural
images. The averaged AUROC results over 50 images are reported in
table~\ref{table9} . This table confirms the higher performance of
the proposed method ({\small CT-GARCH}) in comparison with {\small
CT-GG} and {\small WT-GARCH}.

\begin{table}
\begin{center}
\caption{\label{table5} AUROC results under scaling attack
(scaling factor = 0.75) }
\begin{tabular}{c |c |c |c }
\hline
 \hline
  \multicolumn{4} {c} {WDR = -50 dB} \\
 \hline
 \hline
     Image & {\small CT-GG }& {\small WT-GARCH }&  {\small CT-GARCH}   \\
     &   Detector&  Detector&  Detector  \\
\hline
  Peppers &0.7877&0.7516&0.9920\\
  \hline
  Living room&0.7578&0.8335&0.9986\\
  \hline
  Lake&0.7510&0.8532&0.9941 \\
  \hline
 Pirate &0.8174&0.7870&0.9834\\
   \hline
\hline
\multicolumn{4} {c} {WDR = -45 dB} \\
 \hline
 \hline
       Peppers &0.8431&0.9026&1.0000\\
  \hline
  Living room&0.8748&0.9115&1.0000\\
  \hline
  Lake&0.8334&0.9319&1.0000 \\
  \hline
 Pirate&0.8873&0.9128&1.0000\\
   \hline
\hline
\end{tabular}
\end{center}
\end{table}

\begin{table}
\begin{center}
\caption{\label{table6} AUROC results under median filtering
attack (window size = $5 \times 5$)
 }
\begin{tabular}{c |c |c |c }
\hline
 \hline
  \multicolumn{4} {c} {WDR = -50 dB} \\
 \hline
 \hline
       Peppers &0.8237&0.8037&0.9971\\
  \hline
  Living room&0.9994&0.8820&1.0000\\
  \hline
  Lake&0.9959&0.7193&0.9996\\
  \hline
 Pirate&0.6258&0.9964&0.8784\\
   \hline
\hline
\multicolumn{4} {c} {WDR = -45 dB} \\
 \hline
 \hline
       Peppers &1.0000&0.8965&1.0000\\
  \hline
  Living room&1.0000&0.9676&1.0000\\
  \hline
  Lake&1.0000&0.9140&1.0000\\
  \hline
 Pirate&1.0000&0.9742&1.0000\\
   \hline
\hline
\end{tabular}
\end{center}
\end{table}

\begin{table}
\begin{center}
\caption{\label{table7} AUROC results under Gaussian filtering
attack (window size = $5 \times 5$)
 }
\begin{tabular}{c |c |c |c }
\hline
 \hline
  \multicolumn{4} {c} {WDR = -60 dB} \\
 \hline
 \hline
     Image & {\small CT-GG }& {\small WT-GARCH }&  {\small CT-GARCH}   \\
     &   Detector&  Detector&  Detector  \\
\hline
  Peppers &0.7000&0.6469&0.9150\\
  \hline
  Living room&0.5182&0.6264&0.9541\\
  \hline
  Lake&0.7962&0.6249&0.8556\\
  \hline
 Pirate &0.8108&0.6622&0.8904\\
   \hline
\hline
\multicolumn{4} {c} {WDR = -50 dB} \\
 \hline
 \hline
       Peppers &0.9439&0.8565&1.0000\\
  \hline
  Living room&0.9269&0.8808&1.0000\\
  \hline
  Lake&0.9152&0.8902&1.0000\\
  \hline
 Pirate&0.9965&0.8795&1.0000\\
   \hline
\hline
\multicolumn{4} {c} {WDR = -45 dB} \\
 \hline
 \hline
       Peppers &{ 0.7300}&0.9700&1.0000\\
  \hline
  Living room&0.9993&0.9789&1.0000\\
  \hline
  Lake&0.9994&0.9362&1.0000\\
  \hline
 Pirate&1.0000&0.9816&1.0000\\
   \hline
\hline
\end{tabular}
\end{center}
\end{table}

\begin{table}
\begin{center}
\caption{\label{table8} AUROC results under combination of
Gaussian Filtering  and AWGN attacks }
\begin{tabular}{c |c |c |c }
   \hline
\hline
\multicolumn{4} {c} {WDR = -50 dB} \\
 \hline
 \hline
 Image & {\small CT-GG }& {\small WT-GARCH }&  {\small CT-GARCH}   \\
     &   Detector&  Detector&  Detector  \\
     \hline
       Peppers &0.8864&0.7773&0.9649\\
  \hline
  Living room&0.8299&0.7268&0.9936\\
  \hline
  Lake&0.8732&0.8195&0.9761\\
  \hline
 Pirate&0.6160&0.7090&0.9772\\
   \hline
\hline
\multicolumn{4} {c} {WDR = -45 dB} \\
 \hline
 \hline
        Peppers &0.7604&0.8960&1.0000\\
  \hline
  Living room&0.9556&0.9084&1.0000\\
  \hline
  Lake&0.5300&0.9143&1.0000\\
  \hline
 Pirate&0.8686&0.8887&1.0000\\
   \hline
\hline
\end{tabular}
\end{center}
\end{table}
%

\begin{table}
\begin{center}
\caption{\label{table9} Average AUROC results for 50 natural
images under different types of attacks (WDR = -50 $dB$) }
\begin{tabular}{c |c |c |c }
\hline
 \hline

     Attack & {\small CT-GG }& {\small WT-GARCH }&  {\small CT-GARCH}   \\
   type  &   Detector&  Detector&  Detector  \\
\hline
  Compression  &0.8463&0.8454&0.9320\\
  (QF=60)&&&\\
  \hline
  Scaling  &0.7863& 0.7982& 0.9654\\
  (SF=0.75)&&&\\
  \hline
 Median Filtering&0.9324&0.8864&0.9828\\
 $5 \times 5$&&&\\
  \hline
Gaussian Filtering&0.9288&0.8705&0.9951\\
$5 \times 5$&&&\\
\hline
 Gaussian Filtering&0.7536&0.7585&0.9708\\
 + AWGN &&&\\
   \hline
\hline
\end{tabular}
\end{center}
\end{table}

\section{conclusion}
This paper  proposed a novel watermark detector for contourlet
domain additive image watermarking. Watermark detection can be
formulated as a binary hypothesis test. Based on Neyman-Pearson
criterion, the optimal detector can be achieved using LRT.
Selecting the statistical model employed in the LRT is a major
issue.
\\
In this paper \cite{item1}, we have first studied the statistical
characteristics of the contourlet coefficients. The
heteroscedasticity of these coefficients has been shown  using two
Lagrange multiplier tests. The previously proposed statistical
models usually assume the contourlet coefficients to be
identically distributed and cannot capture the heteroscedasticity
of these coefficients. So, these models don't provide a good
compatibility with the contourlet coefficients and the watermark
detectors based on such models show inadequate performance. To
overcome this problem, we studied the compatibility between the
contourlet coefficients and 2D-GARCH which is an efficient and
flexible heteroscedastic model. Consequently, we designed a
watermark detector based on 2D-GARCH model. Our method is based on
solid statistical theory and the ROC of 2D-GARCH based detector
has been derived analytically. We have studied the experimental
efficiency of the proposed detector in detail by conducting
several experiments. The robustness of the proposed detector
against different kinds of attacks has been evaluated and the
superiority of the proposed method compared with some other
methods has been shown \cite{item18,item1}.\\
This paper is the first work that studies and captures the heteroscedasticity  of the contourlet coefficients, and taking into account the heteroscedasticity of the contourlet coefficients can lead to good results in other fields such as contourlet domain image restoration.

\appendix
\section{Appendix A}
Here we compute $\mu_0,\sigma_0,\mu_1,$ and $\sigma_1$ as defined in section 5.1. To simplify the notation, we define
two functions: {\small:
\begin{eqnarray}
\nonumber \mathcal{X}({\bf F}_{ij},\gamma)=
\quad \quad \quad \quad \quad \quad \quad \quad \quad \quad
\quad \quad \quad \quad \quad \quad \quad \quad \quad \quad \quad \quad \quad\\
\nonumber \sqrt{\frac{\sum_{k\ell \in \Omega _1 } {\alpha _{k\ell}
f_{i - k,j - \ell}^2}  + \sum_{k\ell \in \Omega _2 } {\beta
_{k\ell} h_{i - k,j - \ell}^{} }}{\sum_{k\ell \in \Omega _1 }
{\alpha _{k\ell} (f_{i - k,j - \ell}-\gamma)^2 }  + \sum_{k\ell
\in \Omega _2 } {\beta _{k\ell} h_{i - k,j - \ell}^{} }}} \quad
\quad \quad \quad  \\ \nonumber \mathcal{Y}({\bf F}_{ij},\gamma)=
\quad \quad \quad \quad \quad \quad \quad \quad \quad \quad \quad
\quad \quad \quad \quad \quad \quad \quad \quad \quad \quad \quad
\quad \\ \nonumber 0.5(\frac{{  {f}_{ij} ^2 }}{{2\alpha _0  +
\sum_{k\ell \in \Omega _1 } {\alpha _{k\ell} f_{i - k,j - \ell}^2
}  + \sum_{k\ell \in \Omega _2 } {\beta _{k\ell} h_{i - k,j -
\ell}^{} } }}) \quad \quad \quad \\+\frac{{ - ({f}_{ij}-\gamma) ^2
}}{{2\alpha _0  + \sum_{k\ell \in \Omega _1 } {\alpha _{k\ell}
(f_{i - k,j - \ell}-\gamma)^2 }  + \sum_{k\ell \in \Omega_2 }
{\beta _{k\ell} h_{i - k,j - \ell}^{} } }} ) \quad
\end{eqnarray} }
where $\Omega_1,\Omega_2,h_{i,j}$ are as defined in section 3.1
and ${\bf F}_{ij}=\{f_{n,m}|1 \leq n \leq i, 1 \leq m \leq j  \}$.
The watermark sequence is an i.i.d. two dimensional random process
which takes the values $+\gamma$ or $+\gamma$ with the same
probability. In this case, using (\ref{e5}),  the mean of $\log
\Lambda$ under the condition $\mathcal{H}_0$ can be computed as
{\small
\begin{eqnarray}
\nonumber \mu_0&=& E_w(\log
\Lambda|\mathcal{H}_0)=0.5(\sum_{i,j}{\log \mathcal{X}({\bf F}_{ij},\gamma) +\log \mathcal{X}({\bf F}_{ij},- \gamma)} )\\
\label{e66} &+& 0.5(\sum_{i,j}{ \mathcal{Y}({\bf F}_{ij},\gamma) +
\mathcal{Y}({\bf F}_{ij},- \gamma)} )
\end{eqnarray}}
where $E_w(.)$ denotes the expectation on $w$. To calculate the
variance of $\log \Lambda$ conditioned on $\mathcal{H}_0$, first
{\small $E_w(\log \Lambda|\mathcal{H}_0)^2$} is computed: {\small
\begin{eqnarray}
\nonumber &E&_w(\log \Lambda|\mathcal{H}_0)^2=E_w(\sum_{i,j} \log \mathcal{X}({\bf F}_{ij},w)+ \sum_{i,j} \mathcal{Y}({\bf F}_{ij},w))^2\\
\nonumber &=&0.5(\sum_{i,j} \log \mathcal{X}({\bf F}_{ij},\gamma)+
\sum_{i,j} \mathcal{Y}({\bf F}_{ij},\gamma))^2\\ \label{e77}
&+&(\sum_{i,j} \log \mathcal{X}({\bf F}_{ij},-\gamma)+ \sum_{i,j}
\mathcal{Y}({\bf F}_{ij},-\gamma))^2
\end{eqnarray}}
then, we have {\small
\begin{eqnarray}
\label{e88} \sigma_0^2=E_w(\log \Lambda|\mathcal{H}_0)^2-\mu_0
\end{eqnarray}}
In the same way, we can compute the mean  of $\log \Lambda$ conditioned on $\mathcal{H}_1$ as
{\small
\begin{eqnarray}
\nonumber \mu_1&=& E_w(\log
\Omega|\mathcal{H}_1)=0.5(\sum_{i,j}{\log \mathcal{X}^{-1}({\bf F}_{ij},\gamma) +\log \mathcal{X}^{-1}({\bf F}_{ij},- \gamma)} )\\
\label{e99} &+& 0.5(\sum_{i,j}{- \mathcal{Y}({\bf F}_{ij},\gamma)
- \mathcal{Y}({\bf F}_{ij},- \gamma)} ).
\end{eqnarray}
To compute the variance of $\log \Lambda$ conditioned on $\mathcal{H}_1$, we have
\begin{eqnarray}
\nonumber E_w(\log \Omega|\mathcal{H}_1)^2 &=& E_w(\sum_{i,j} \log \mathcal{X}^{-1}({\bf F}_{ij},w)+ \sum_{i,j} -\mathcal{Y}({\bf F}_{ij},w))^2\\
\nonumber &=&0.5(\sum_{i,j} \log \mathcal{X}^{-1}({\bf F}_{ij},\gamma)+ \sum_{i,j} -\mathcal{Y}({\bf F}_{ij},\gamma))^2\\
\label{e101} &+&(\sum_{i,j} \log \mathcal{X}^{-1}({\bf
F}_{ij},-\gamma)+ \sum_{i,j} -\mathcal{Y}({\bf F}_{ij},-\gamma))^2
\end{eqnarray}
and
\begin{eqnarray}
\label{e111} \sigma_1^2 &=& E_w(\log \Omega|\mathcal{H}_1)^2-\mu_1
\end{eqnarray}}




\end{document}